\documentclass[journal,draftcls,onecolumn,12pt,twoside]{IEEEtranTCOM}
\normalsize
\renewcommand{\baselinestretch}{1.5}

\usepackage{amsmath,amssymb}
\usepackage{times}
\usepackage{relsize}
\usepackage{graphicx}
\usepackage{epstopdf}
\usepackage{url}
\usepackage{booktabs}
\usepackage{multirow}
\usepackage{mparhack}
\usepackage{subfigure}
\usepackage{authblk}
\usepackage{amsthm}

\usepackage{algorithm}
\usepackage{algorithmicx}
\usepackage{algpseudocode}

\usepackage{array}
\usepackage{cite}
\usepackage{eqparbox}
\usepackage{mdwmath}
\usepackage{balance}
\usepackage{epsfig}
\usepackage{xcolor}

\usepackage{mathtools}
\usepackage{bm}
\usepackage{amsfonts}
\usepackage[all]{xy}
\usepackage{etoolbox}
\usepackage{graphicx}
\usepackage{setspace}
\usepackage{caption}

\DeclareMathAlphabet{\mathantt}{OT1}{antt}{li}{it}
\DeclareMathAlphabet{\mathpzc}{OT1}{pzc}{m}{it}

\usepackage{accents}

{%

\makeatletter

\makeatother

\ifCLASSINFOpdf
\else
\fi

\hyphenation{op-tical net-works semi-conduc-tor}

\begin{document}
%
\title{ Grant-Free Transmission by LDPC Matrix Mapping and Integrated Cover-MPA Detector
}
%
%
%

\author{Linjie~Yang,~\IEEEmembership{Student Member,~IEEE},
	Pingzhi~Fan,~\IEEEmembership{Fellow,~IEEE}, Li~Li,~\IEEEmembership{Member,~IEEE}, 
	Zhiguo~Ding,~\IEEEmembership{Fellow,~IEEE}, Li~Hao,~\IEEEmembership{Member,~IEEE}
	\thanks{Linjie Yang, Pingzhi Fan, Li Li, and Li Hao are with the School of
		Information Science and Technology, Southwest Jiaotong University, Chengdu
		610031, China (e-mail: yanglinjie@my.swjtu.edu.cn; pzfan@swjtu.edu.cn;
		ll5e08@home.swjtu.edu.cn; lhao@swjtu.edu.cn).}
	\thanks{Zhiguo Ding is with the Department of Electrical Engineering, Princeton
		University, Princeton, NJ 08544 USA, and also with the Department of Electrical
		and Electronic Engineering, The University of Manchester, Manchester
		M13 9PL, U.K. (e-mail: zhiguo.ding@manchester.ac.uk).}
}

\markboth{IEEE Transactions on Communications}%
{Submitted paper}
%



\maketitle

\begin{abstract}
	In this paper, a novel transceiver architecture is proposed to simultaneously achieve
	efficient random access and reliable data transmission in massive IoT networks. 
	At the transmitter side, each user is assigned a unique protocol sequence 
	which is used to identify the user and also indicate
	the user's channel access pattern.	
	Hence,  user identification is completed by
	the detection of channel access patterns. 
	Particularly, the columns of a parity check matrix {of low-density-parity-check (LDPC) code}  are employed as  protocol sequences. 
	The design guideline of this LDPC parity check matrix and the associated performance analysis are provided in this paper.
	At the receiver side, a two-stage iterative detection architecture is designed, which consists of a group testing component and a payload data decoding component. They collaborate in a way that the group testing component maps detected protocol sequences to a tanner graph, on which the second component could execute its message passing algorithm. In turn, zero symbols  detected by {the}  message passing algorithm of the second component indicate  potential false alarms made by the first group testing component. Hence, the tanner graph could iteratively evolve.
	The {provided simulation} results demonstrate that our transceiver design realizes a practical one-step grant-free transmission and has a compelling performance.
\end{abstract}
\begin{IEEEkeywords}
	Grant-Free, Short-Packet-Transmission, Group Testing, Cover Decoder, Iterative Detection. 
\end{IEEEkeywords}

\IEEEpeerreviewmaketitle

\section{introduction}
Massive machine type communication (mMTC)  has attracted 
comprehensive research interests in recent years, since it was identified as one of the three major services of the fifth generation mobile communication (5G) system.  Especially, after the freeze of 3GPP release-16 \cite{au2020short},
{the} requirements of enhanced mobile broadband communications (eMBB) had been well satisfied. Hence, researchers turn to dedicate  efforts to solve the challenges for the implementation of mMTC.
The mMTC service has its particular features:
(1) compared with human-to-human (H2H) communications, the number of potential users (devices) in
mMTC scenario could reach up to millions {per $\rm km^{2}$} \cite{r11808866}; (2) the user activity patterns are very sporadic; (3) mMTC
demonstrates a salient short-packet-transmission property \cite{bockelmann2018towards}; (4) the users are very sensitive to energy
consumption \cite{r11905187}. 
As a result, employing	
traditional
four-step random access in mMTC scenarios will induce a {challenging} overhead problem that typically, hundreds of signaling bits are transmitted for facilitating the transmission of just several payload bits. 
Among {the} solutions of this high overhead problem, grant-free (GF) random access (RA) is a promising candidate
which aims at implementing the synchronization, active user identification, channel estimation, as well as the data recovery in a one-shot joint operation \cite{ieee2007ieee}.
Currently, compressive sensing (CS) based GF RA \cite{chen2018sparse,liu2018massive,liu2018sparse}, 
unsourced random access (URA) \cite{polyanskiy2017perspective,ordentlich2017low,pradhan2020polar,pradhan2019joint,zheng2020polar} and GF non-orthogonal multiple access (GF-NOMA) \cite{oyerinde2019compressive,yu2020binary,liu2020joint,adnanimproved,abed2021cs} constitute main GF techniques.

Most of the existing CS based GF RA schemes leverage the particular feature of mMTC users that they normally access the channel in a sporadic manner. Hence, the active user identification and channel estimation engaged in RA process could be accomplished by CS technologies. In more depth, depending on the number of antennas deployed at the BS, they could be formulated to either a single measurement vector problem \cite{chen2018sparse} or a multiple measurement vector problem \cite{liu2018sparse}. It was also demonstrated in \cite{chen2018sparse} and \cite{liu2018massive} that {the} approximate message passing {(AMP)} algorithm contributes an efficient solution of these problems. While further considering the data transmission, Larrson \emph{et al}. developed the prototype in \cite{chen2018sparse,liu2018massive} by embedding user data bits directly in its optional random access preambles. Hence, a one-step GF transmission framework is realized in \cite{liu2018sparse}. But, the number of data bits that could be afforded by this framework is quite small, e.g., several bits per user. Hence, {designing} a practical one-step GF transmission framework is still {an important open problem}.

URA technologies pave another way to approaching one-step GF transmission. Polyanskiy \emph{et al}. firstly proposed the URA concept  \cite{polyanskiy2017perspective} by regarding the active user identification and data recovery as a fundamental coding problem. In \cite{polyanskiy2017perspective}, all the users employ the same codebook, {where user} identity is regarded as a part of entire data bits
{and the}
theoretical performance bound of URA framework is derived. Polyanskiy \emph{et al}. also exhibited a practical realization of {the} framework in  \cite{ordentlich2017low}, which was referred as T-fold ALOHA. A two-stage receiver consisting of outer and inner coding components plays a critical role in T-fold ALOHA. 
{A two-stage receiver architecture was employed in ensuing developments of URA technology}
\cite{pradhan2020polar,pradhan2019joint,zheng2020polar}, where the first stage {is for} user identification while the second stage {is for data} recovery. Compared with \cite{ordentlich2017low}, more sophisticated anti-jamming techniques were proposed for GF in \cite{pradhan2020polar,pradhan2019joint,zheng2020polar}.

The development of NOMA technology paralleled the explosion of IoT devices, since it can support severely overloaded communication, where the number of active users significantly exceeds that of channel resources. After abandoning the idealized assumption that BS always has a prior knowledge of user-activity, GF NOMA technologies occurred. Accordingly, they focused on solving active user identification before implementing non-orthogonal multiple access. Dynamic compressive sensing, subspace pursuit as well as orthogonal matching pursuit algorithms were {developed} in \cite{oyerinde2019compressive}; Golay complementary sequence is employed as a new kind of preamble in  \cite{yu2020binary}; a two-stage receiver realizing joint user-activity and data detection is designed for NOMA uplink in \cite{liu2020joint}.

The mechanism that an active user needs to send a user-specific preamble to facilitate its identification at BS is adopted in most of the aforementioned GF RA frameworks. Benefiting from the intrinsic sparsity of the massive IoT network and leveraging the advanced CS technologies, the preamble length required by them has been
significantly less than that required by conventional orthogonal RA under the constraint of achieving the same RA success probability.
However, normally, the preamble is till extremely longer than data stream {due to} the short-packet-transmission property of IoT users.

In order to essentially solve the overhead problem caused by directly sending preambles 
to distinguish users, the literatures \cite{yu2019multiuser,li2021efficient,8849823} alternatively embed user identity in its channel access pattern. 
To be {more} concrete, the same data packet of a user will be repeated on multiple channel resources. The distribution of these loaded channel resources is termed {the} user channel access pattern. Hence, it is feasible to establish one-to-one mapping  between  {the user} identity and channel access pattern. Furthermore, the channel access pattern of active users can be detected by {the} group testing (GT) algorithm \cite{8849823}. However, currently, the performance of group testing is still constrained by the active user sparsity. {It works well only in extremely sparse scenario}. How to coordinate the channel access patterns of all the coexisted users for supporting more simultaneously activated users is still an important open problem \cite{zahidah2018construction,sasmal2019disjunct}. In \cite{zhu2010exploiting}, a more radical solution was proposed, which even did not have an explicit active user identification stage. Its message passing algorithm based receiver directly decodes the transmitted symbol of every coexisted users. Particularly, it adds zero {symbols} to the search space. If the transmitted symbol of a user is decoded {as} zero, it will be regarded as an inactive user.
Apparently, the complexity of this solution rapidly becomes prohibitive upon increasing the number of coexisted users.
Furthermore, in \cite{zhang2020bayesian}, a novel GF low-density-signature orthogonal frequency division multiplexing (GF LDS-OFDM) is proposed, where its joint active user identification and channel estimation (CE) is realized without using preamble. In contrast, its joint active user identification and CE is formulated as an estimation problem of structural signals, and message passing based Bayesian algorithms are employed to solve this estimation problem. 
	However, it does not has any information exchange mechanism to further justify
or correct the user identification results. Again, unfortunately, its	 performance will degrade dramatically with the growth of user sparsity.

In order to  improve the above-stated one-step GF short-packet-transmission, which avoids  explicit preamble transmissions, we propose a novel  transceiver architecture, whose main contributions are summarized as follows.

\begin{itemize}
	
	\item  {A novel GF short-packet transmission scheme is proposed, where the active user identification and data decoding is realized jointly without using preamble.} At the transmitter side, inspired by \cite{8849823}, the user identity is bound to its channel access pattern, which is indicated by a binary protocol sequence. A coordinated construction of the overall protocol sequence set is realized by mapping columns of a proper LDPC parity check matrix to {the} protocol sequences.

	\item At the receiver side, a two-stage iterative detection architecture is designed, which consists of a group testing component and a payload data decoding component. The first-stage group testing component maps its detected protocol sequences to a tanner graph, on which the second-stage data decoding component could execute its message passing algorithm. In the initial iteration, the cover decoder serves as our group testing algorithm to overcome the absence of {the} prior knowledge from {the} data decoding component. {After the initial iteration}, the cover decoder is {replaced iteratively} by a simple belief propagation (BP) algorithm, which converts zero symbol probabilities fed back from {the} data decoding component to {the} user activity states. Hence {the} potentially redundant nodes could be pruned from previous tanner graph used in {the} data decoder. In this manner, the receiver performance could be iteratively improved.
	
	\item 
	The design guideline of the aforementioned LDPC parity-check-matrix based user specific protocol sequences is provided, which appropriately couples  
	with the decoding rule of cover decoder that employed at the first detection stage. 
	Accordingly, the optimization of its LDPC parity-check-matrix is discussed. Moreover,
    the computational complexity of the proposed two-stage iterative detection algorithm is also analyzed.

\end{itemize}

The rest of this paper is organized as follows. The system model is described in Section II. The work of the group testing component and the data decoding component are discussed in Section III and Section IV respectively. Simulation results is presented in Section V. The paper is concluded in Section VI.

$Notations:$ In this paper, $\mathbf{S}^{H}$ denotes the conjugate transpose  of  matrix $\mathbf{S}$.
$\mathbf{S}[:,\mathcal{U}]$ consists of a set of columns in $\mathbf{S}$, whose indices are given in $\mathcal{U}$. $|\mathcal{A}|$ denotes the number of elements in the set $\mathcal{A}$. $||\cdot||_{p}$ denotes the $l_{p}$-norm.
\begin{figure}[htbp]
	\centering
	\includegraphics[scale=1.2]{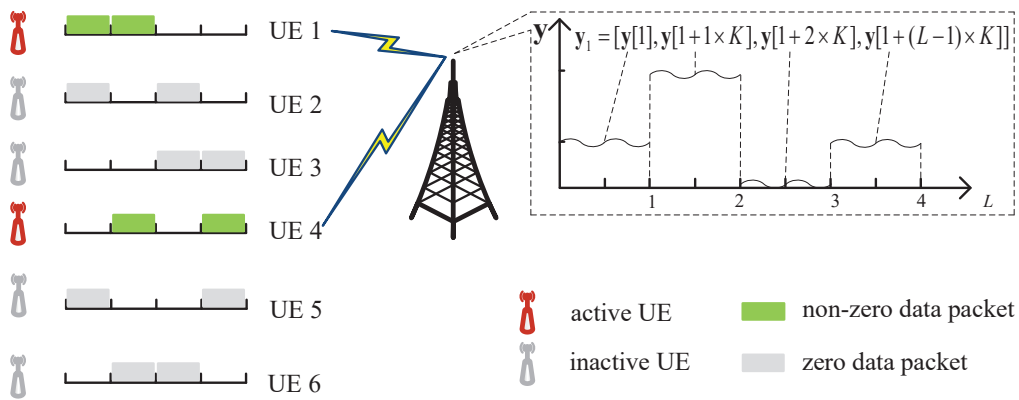}
	\caption{The proposed one-step grant-free uplink transmission framework.}
	\label{system_model}
\end{figure}

\section{System Model}
In this section, the system model of our one-step GF short-packet-transmission is elaborated. {The} important assumptions adopted in our system are emphasized {first}.

We focus on the uplink of a single IoT cell. We assume that a number of $N$ users coexist in the cell, among {which}, a number of $N_{a}$ users will independently wake up and send data in once GF transmission period. Accordingly, the cellular user sparsity is {defined as} $\lambda = \frac{N_{a}}{N}$. During {one} random access procedure, a total number of $L$ channel resources are available for carrying all uplink transmissions of active users. Without loss of generality,  $L$ uniform time-slots are considered in this paper and each of them has the same length as a data-packet \footnote{\ Alternatively, they could be $L$ uniform sub-bands in the frequency domain.}. For simplifying the signaling and analysis, we further assume the system always achieves an idealized symbol-wise synchronization and experiences AWGN channels. Normally, the IoT {devices are low-cost and energy constrained, which motivates the used assumption that a single antenna is deployed in both the user and the BS.}

Each cellular user $u$ is {assigned} with a unique protocol sequence $\mathbf{s}_{u}\in \mathbb{B}^{L\times 1}, u=1,2,\cdots,N$.
As addressed in Section I, $\mathbf{s}_{u}$ can also be termed as the channel access pattern of user $u$.
A user $u$ generates its original alphabet {$\mathcal{X}_{u}$}  according to \cite{van2009multiple}, where {each} standard $M$-ary phase shit keying (PSK) symbol is further multiplied by a user-specific complex coefficient. Then, similar to the sparse CDMA system discussed in \cite{zhu2010exploiting}, an extra zero constellation point is added to  its original  alphabet for exposing its inactive state at BS. Hence, the final alphabet of user $u$ has a cardinality of $\vert \mathcal{X}_{u} \vert = \mathcal{M}+1, u=1,2,\cdots, N$.

When the one-step GF transmission of a user $u$ commences, it firstly switches its activity state from $a_{u}=0$ to $a_{u}=1$. Then, its data packet {of length $K$,} $\mathbf{x}_{u}=[ x_{u}[1], x_{u}[2] ,\cdots,x_{u}[K] ]^{T} \in \mathbb{C}^{K\times 1}$ is prepared, whose component symbols $x_{u}[k], k=1,2,\cdots, K$, are selected from $\mathcal{X}_{u}$.
Subsequently, the data packet $\mathbf{x}_{u}$ is spread by its protocol sequence $\mathbf{s}_{u}$ in the form of $\mathbf{x}_{u}^{c} = \mathbf{s}_{u}\bigotimes\mathbf{x}_{u} \in \mathbb{C}^{KL\times 1}$, where $\bigotimes$ denotes the Kronecker product and $\mathbf{x}_{u}^{c}$  {denotes} the spread data-packets of user $u$.

Assuming all the active users are perfectly aligned in time, the signal sequence received at the BS during {a single} GF transmission period is given by 
\begin{equation}
\mathbf{y} = \sum_{u=1}^{N} \mathbf{x}_{u}^{c}\cdot a_{u} + \mathbf{n}.
\end{equation} 
Then, by sampling $\mathbf{y}$ at $k^{th}$ symbol duration of every time-slot and assembling these $L$ samples, we glean the $k^{th}$ sub-vector of $\mathbf{y}$, which could be formulated as
\begin{equation}
\begin{aligned}
\mathbf{y}_{k} &= \sum_{u=1}^{N}  a_{u}[\mathbf{x}_{u}^{c}[k],\mathbf{x}_{u}^{c}[k+ K\cdot 1],\cdots, \mathbf{x}_{u}^{c}[k+K(L-1)]  ]^{T} + \mathbf{n}_{k} \\
&= \mathbf{S}\mathbf{X}_{k}\mathbf{a} + \mathbf{n}_{k},
\end{aligned}
\end{equation}
where $\mathbf{S}=[\mathbf{s}_{1}, \mathbf{s}_{2}, \cdots, \mathbf{s}_{N}]\in \mathbb{B}^{L\times N}$. $\mathbf{X}_{k}= \rm{diag}(x_{1}[k], x_{2}[k],\cdots, x_{N}[k]) \in \mathbb{C}^{N\times N}$, $\mathbf{a}=[a_{1},a_{2},\cdots,\\ a_{N}]^{T}$, and $\mathbf{n}_{k}$ is an {$L\times 1$} Gaussian vector, whose element {follows} an i.i.d distribution of $\mathcal{CN}(0,\delta^{2})$.

An  example of our {considered} one-step GF uplink transmission framework is illustrated in Fig. \ref{system_model}, where $N=6$ IoT users coexist in the cell and $L=4$ time-slots are employed for {a single} GF transmission period. Hence, the length of protocol sequence $\mathbf{s}_{u}, u=1,\cdots,6$ is also fixed to $L=4$. On the left-hand side of Fig. \ref{system_model}, $\mathbf{s}_{u}$ is directly visualized by channel access pattern of user $u$, whose non-zero element indicates at which time-slot a data packet should be loaded, e.g. $\mathbf{s}_{1}=[1, 1, 0, 0]$. 
{Assume a pair of symbols are delivered in one time slot}, i.e. $K=2$, the data packet of user 1 could be formulated as $\mathbf{x}_{1}=[x_{1}[1], x_{1}[2]]^{T}$. After spreading $\mathbf{x}_{1}$ by $\mathbf{s}_{1}$, user 1 sends its entire signal frame
$\mathbf{x}^{c}_{1}=[ x_{1}[1], x_{1}[2],x_{1}[1], x_{1}[2],0,0,0,0  ]^{T}$. Assume {that} user 1 and 4 are active, other users are asleep, according to  (2), the $1^{st}$ sub-vector of received signal sequence is given by
\begin{small}
	\begin{equation}
	\left[
	\begin{matrix}
	y[1] \\
	y[3] \\
	y[5] \\
	y[7] 
	\end{matrix}
	\right]
	= 
	\left[
	\begin{matrix}
	1 & 1  &   &0 \\
	1 & 0  &\cdots  &1 \\
	0 & 1  &  &1 \\
	0 & 0  &  &0 \\
	\end{matrix}
	\right]\cdot
	\left[
	\begin{matrix}
	x_{1}[1] &  &  & \\
	&x_{2}[1]  &  & \\
	&  &\ddots  & \\
	&  &  &x_{6}[1] \\
	\end{matrix}
	\right]\cdot
	\left[
	\begin{matrix}
	1\\
	0\\
	0\\
	1\\
	0\\
	0\\
	\end{matrix}
	\right] + \mathbf{n}_{1},
	\end{equation}
\end{small}
\begin{figure*}[htbp]
	\centering
	\includegraphics[scale=1.2]{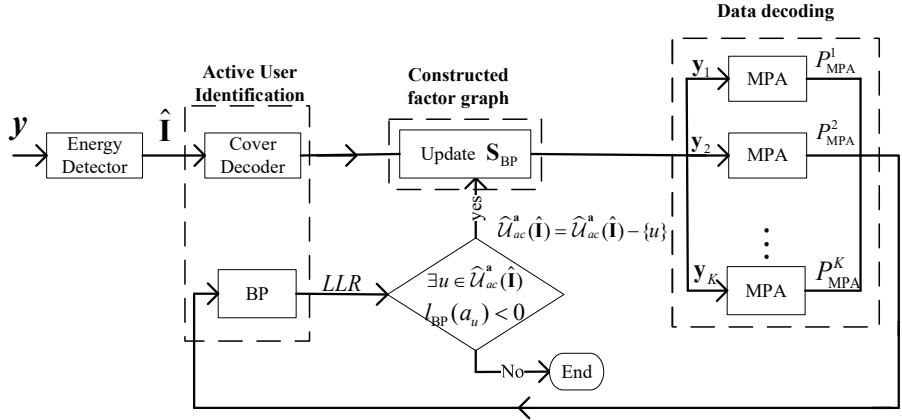}
	\caption{Schematic of the proposed two-stage iterative active user identification and data recovery.}
	\label{SIC-MPA}
\end{figure*}

\section{Active User Identification }
{The} Message passing algorithm (MPA) {was} leveraged in \cite{hoshyar2008novel} to recover multi-user symbols from 
{the}
superposed signals, e.g. from $\mathbf{y}_{k}$ given in (2). Its {key idea} is representing the channel access patterns of all  users in a form of {the} LDPC parity-check matrix, e.g. the matrix $\mathbf{S}$ in (2). 
{Hence, the factor graph required by the MPA algorithm is obtained according to $\mathbf{S}$.}
This scheme was adopted as a fundamental technology in many NOMA systems \cite{chen2016pattern,zhang2014sparse}. However, it {cannot} be directly applied to our case owing to {the fact that} $N$ is normally quite large in mMTC scenario. The extremely high dimensionality of $\mathbf{S}$ will impose  prohibitive complexity on MPA. Hence, we have to implement active user identification before data recovery. Accordingly, the 
columns\footnote{{\ $\mathcal{U}_{ina}$ }is the set of asleep users.} $\mathbf{s}_{n}, n\in \mathcal{U}_{ina}$ can be removed from $\mathbf{S}$. This pruned $\mathbf{S}$ will significantly reduce the computing complexity of MPA and mitigate the distortion caused by false-alarmed users as well.

Inspired by \cite{8849823}, again, active user identification is regarded as a group testing problem in this paper. But its performance is improved by us through two aspects: a) instead of disjunct matrices \cite{zahidah2018construction,sasmal2019disjunct}, the protocol sequence used in {the} cover decoder which is obtained from a properly constructed LDPC parity-check matrix; b) not only {the} cover decoder but also {the} belief propagation (BP) decoder are employed in group testing procedure, where the later enable the receiver to iteratively update $\mathbf{S}$ based on the feedbacks from data decoder. 


\subsection{General Principle of Cover Decoder}
Let $\mathbf{I}$ denote the load states of {each} time-slot during {one} GF transmission.
If data packets of any active user are transmitted on $l^{th}$ time-slot, we have $\mathbf{I}[l]=1$, otherwise, $\mathbf{I}[l]=0$. According to these definitions, $\mathbf{I}$ is determined by
\begin{equation}
\mathbf{ I } = \mathop{\vee} \limits_{u \in  \mathcal{U}_{ac}^{\mathbf{a}}  } \mathbf{s}_{u},
\end{equation}
where   
$\mathcal{U}_{ac}^{\mathbf{a}}=\{u\vert \mathbf{a}[u]=1,1\leq u\leq N\}$
indicates the set of active users 
and $\vee$ denotes Boolean-OR operation of 
{vectors. Obviously, $\mathcal{U}^{\mathbf{a}}_{ac}$ is determined by its superscript $\mathbf{ a }$, which is defined in (2). }

The cover decoder completes its active user identification through three steps:
\begin{itemize}
	\item Firstly, load state $\mathbf{I }$ is detected as $\hat{\mathbf{ I }}$ by energy detector, which will be addressed in Section II.C.
	\item  {Secondly}, according to the distribution of zero elements in 
	$\hat{\mathbf{ I }}$, the set of inactive users $\mathcal{U}_{ina}^{\mathbf{a}}$ can be estimated.
	\item Finally, {assuming that} $\mathcal{U}_{ina}^{\mathbf{a}}$ has been estimated as $\hat{\mathcal{U}}_{ina}^{\mathbf{a}}( \hat{ \mathbf{I} } )$, the set of active users $\mathcal{U}_{ac}^{\mathbf{a}}$ is simply estimated by its complementary set of 
	$   \hat{ \mathcal{U} }_{ac}^{\mathbf{a} }( \hat{\mathbf{ I }} ) =  \mathcal{U} - \hat{\mathcal{U}}_{ina}^{\mathbf{a}}( \hat{ \mathbf{I} } )$.
	
\end{itemize}

Take Fig. \ref{system_model} as an example again. According to the channel load states depicted on the top-right corner of Fig. \ref{system_model}, we 
assume perfect load state estimation, i.e. $\hat{\mathbf{I}}=\mathbf{ I }=[ 1, 1, 0, 1  ]$. The zero element on the third index of $\mathbf{I}$ determines {whether it is possible} that a user whose channel access pattern occupies the third time-slot is an active user during current GF transmission. Hence, the set of inactive users is estimated as
$\hat{\mathcal{U}}_{ina}^{\mathbf{a}}( { \mathbf{I} } )=\{2, 3, 6\}$.
Accordingly, the set of active users is estimated as 
$\hat{ \mathcal{U} }_{ac}^{\mathbf{a} }( {\mathbf{ I }} )=\{1, 4, 5\}$.
Obviously, in this case,  
$\hat{ \mathcal{U} }_{ac}^{\mathbf{a} }( {\mathbf{ I }} )$
is not completely identical to $\mathcal{U}_{ac}^{\mathbf{a}}$, which does not include user $5$. This problem could be solved by the proposed two-stage iterative detection architecture as shown in Fig. \ref{SIC-MPA}.

Furthermore, based on 
$\hat{ \mathcal{U} }_{ac}^{\mathbf{a} }( \hat{\mathbf{ I }} )$, we can obtain the spreading matrix of active users 
$\mathbf{S}[:, \hat{ \mathcal{U} }_{ac}^{\mathbf{a} }( \hat{\mathbf{ I }} )]$, which could be converted to a factor graph by regarding its columns as variable nodes and rows as check nodes.

\subsection{ Protocol Sequence Design for Cover Decoder}

As suggested in \cite{hoshyar2008novel}, in order to guarantee the efficiency of MPA, 
the users' spreading sequences should be sparse. As a result, the spreading matrix generated in \cite{hoshyar2008novel}, which plays a similar role as our $\mathbf{S}$ in (2), is a sparse matrix and hence may be regarded as {an} LDPC parity check  matrix. This fact inspires us to map an LDPC parity check matrix to our spreading matrix $\mathbf{S}$.
The results in \cite{hoshyar2008novel} show that most LDPC parity check matrices, even those have short cycles, are suitable for {the} LDS scheme. However, in our case, the active user identification performance will demonstrate noticeable variation while assigning different LDPC parity check matrices to $\mathbf{S}$ and using {the} cover decoder. Hence, optimizing {the} LDPC parity check matrix employed in cover decoder algorithm is of great {importance}.

Spreading matrix optimization in the background of cover decoder applications {has} been studied in \cite{zahidah2018construction,sasmal2019disjunct}, where it was also termed {the} non-adaptive group testing (GT) problem. 
Moreover, in these studies, a noiseless case where $\hat{ \mathbf{I} }=\mathbf{ I }$
is assumed in the optimization process of {the} spreading matrix. 
It was proved in \cite{zahidah2018construction,sasmal2019disjunct} that in order to minimize the active user identification error, which could be quantified by 
$\vert \hat{ \mathcal{U} }_{ac}^{\mathbf{a} }( {\mathbf{ I }} ) - \mathcal{U}_{ac}^{\mathbf{a}} \vert $, spreading matrix has to be a disjunct matrix. With the aid of {the} disjunct matrix based design of spreading matrix, error-free active user identification becomes possible. 
However, the penalty is that the user activity pattern of a mobile-cell has to be extremely sparse, i.e. the parameter $\lambda$ has to be extremely small. Otherwise, the active user identification error will dramatically increase.

Proposing such a solution of GT problem in \cite{zahidah2018construction,sasmal2019disjunct} is owing to an underlying constraint that they have to accomplish active user identification in once chance, i.e. only through {the} cover decoder. But now, in our proposed two-stage iterative detection architecture, the false alarm made in {the} cover decoder could be corrected by detecting zero symbols {during} the ensuing data recovery stage. 
Hence, it is no longer necessary to pursuit perfect active user identification of {the} cover decoder.
This relax of $\vert \hat{ \mathcal{U} }_{ac}^{\mathbf{a} }( {\mathbf{ I }} ) - \mathcal{U}_{ac}^{\mathbf{a}} \vert $ {during the} cover decoder stage brings us two benefits: 1) the performance of active user identification can remain {at an acceptable level} in a wider range of $\lambda$; 2) {the} spreading matrix is not necessary to be a disjunct matrix, {which means that} other construction methods, e.g. employing an LDPC parity check matrix, becomes feasible.

Based on above analysis, we propose that the spreading matrix of all coexisted users $\mathbf{S}$ should result in two properties:
\begin{equation}
\mathcal{U}^{ \mathbf{a}  }_{{ac}} \subseteq \hat{\mathcal{U}}_{ac}^{\mathbf{a}}({ \mathbf{I} }),
\end{equation}
\begin{equation}
R_{\rm FA}= \frac{ \sum\limits_{ \mathbf{a} \in \mathbb{B}^{N\times 1}\atop ||\mathbf{a}||_{0}=\lambda N  }  \vert \hat{ \mathcal{U} }_{ac}^{\mathbf{a} }( {\mathbf{ I }} ) - \mathcal{U}_{ac}^{\mathbf{a}} \vert}{ \sum\limits_{\mathbf{a} \in \mathbb{B}^{N\times 1}\atop||\mathbf{a}||_{0}=\lambda N }  |\mathcal{U}^{ \mathbf{a}  }_{{ac}}|} \leq {\tau.}
\end{equation}
where $||\cdot||_{0}$ denotes $l_{0}$-norm. 
Determined by the decoding principle of {the} cover decoder, the {requirement} of (5) is always achievable as long as the load state $\mathbf{ I }$ is perfectly estimated \cite{zahidah2018construction,sasmal2019disjunct}. 
The threshold $\tau$ in (6) specifies the maximum false alarm that {is allowed} in the cover decoder. If $\tau$ is set too large, the data decoding component illustrated in Fig. \ref{SIC-MPA} will suffer from an extremely high decoding complexity. In turn, a very low false alarm level $\tau$ is only possible in a very sparse user access pattern, i.e. the number of active users has to be seriously limited. According to \cite{zhang2014theoretical}, without the \emph{a priori} knowledge of sparsity $\lambda$, the cardinality of 
$\hat{ \mathcal{U} }_{ac}^{\mathbf{a} }({\mathbf{ I }} )$
may exceed twice or even triple of 
$\mathcal{U}_{ac}^{\mathbf{a}}$.
Hence, in order to constrain the tanner graph size {used later by the} MPA algorithm and simultaneously afford more active users, $R_{\rm FA} \leq 1$ is pursued in our design.


\noindent \textbf{Proposition 1.} \emph{By employing a regular LDPC parity check matrix having column weight $w_{c}$, number of rows $L$, number of columns $N$
	and girth larger than 4 as the spreading matrix $\mathbf{S}$. Assume the load states $\mathbf{I}$ is  perfectly estimated. The false alarm performance
	$R_{\rm FA}$  of cover decoder will be a function of $\lambda$, $r$ and $w_{c}$ as follows} 
\begin{equation}
R_{\rm FA}(\lambda,w_{c},r) = \frac{1-\lambda}{\lambda}(1-(1-\lambda)^{\frac{w_{c}}{r}-1})^{w_{c}},
\end{equation}
\begin{proof}
	See Appendix \ref{cover_decoder_performance}.
\end{proof}




It is well known that massive IoT networks are typical energy {constrained} and channel resource limited communication scenarios \cite{ordentlich2017low,8849823}.
Hence, the column weight $w_{c}$, which is equivalent to the repeat times of a data {packet,} should be as small as possible to save transmission energy. In the same spirit, the ratio of $r=\frac{L}{N}$, which is equivalent to the number of channel resources per cellular user required should also be as small as possible. Based on these analysis and 
Proposition 1, optimizing {the} spreading matrix $\mathbf{ S }$ becomes tractable and could be formulated as a multi-objective optimization problem \cite{marler2004survey}.

\begin{equation}
\begin{aligned}
\min \limits_{w_{c},r} \quad &[ w_{c},r ]\\
s.t. \quad &{\rm C_{1}:} \ w_{c}\geq2\\
&{\rm C_{2}:}\  0\textless r \textless 1\\
&{\rm C_{3}:}\  R_{\rm FA}(\lambda^{\star},  w_{c}, r ) \leq \tau 
\end{aligned}
\end{equation}
where $\lambda^{\star}$ is the specific value of $\lambda$, {which maximizes} $R_{\rm FA}$ after $w_{c}$ and $r$ are given.
${\rm C_{1}}$ comes from the constraint of {the} column weight used in regular LDPC parity check matrices. ${\rm C_{2}}$ comes from the fact that we focus on an overloaded communication system, {and} ${\rm C_{3}}$ is added for satisfying (6).

Directly obtain the optimum solution of $w_{c}$ based on its original formulations in (8) is quite difficult. Its major challenge is the monotonicity of $R_{\rm FA}(\lambda^{\star},w_{c},r)$ with respect to $w_{c}$ is unknown. In order to circumvent this problem, the constraint $\rm {C}_{3}$ that $R_{\rm FA}(\lambda^{\star},  w_{c}, r ) \leq \tau $ may be replaced by requiring the upper bound of $R_{\rm FA}(\lambda^{\star},  w_{c}, r )$ to be less or equal than $\tau$. Obviously, this replacement tights previous constraint $\rm {C}_{3}$, but we could prove later that it will not modify the optimum $w_{c}$.

According to the above-mentioned {clarification}, the upper bound {on} $\ln R_{\rm FA}(\lambda,w_{c},r)$ is given by
\begin{equation}
\begin{aligned}
\ln (R_{\rm FA}(\lambda, w_{c}, r)) 
&\leq \ln (\frac{1-\lambda}{\lambda}) - w_{c}(1-\lambda)^{\frac{w_{c}}{r}-1},
\end{aligned}
\end{equation}
where the inequality of $\ln (1-x)\leq -x,x\in(0,1)$ is employed. $R_{\rm FA}(\lambda,w_{c},r)$ is substituted as $1-x$. Let $g(\lambda,w_{c},r)$ represent the upper bound $\ln (\frac{1-\lambda}{\lambda}) - w_{c}(1-\lambda)^{\frac{w_{c}}{r}-1}$. Consequently, constraint $\rm C_{3}$ could be replaced by $g(\lambda,w_{c},r)\leq \ln \tau$. As a result, the optimization problem in (8) is transformed to

\begin{equation}
\begin{aligned}
\min \limits_{w_{c},r} \quad &[ w_{c},r ]\\
s.t. \quad &{\rm C_{1}}\sim {\rm C_{2}},\\
& {\rm C_{4}}: \ g(\lambda^{\star},  w_{c}, r ) \leq \ln \tau 
\end{aligned}
\end{equation}

\noindent \textbf{Proposition 2.} \emph{{By employing} a regular LDPC parity check matrix, having column weight $w_{c}$, number of rows $L$, number of columns $N$  and girth  larger than 4 as the spreading matrix $\mathbf{S}$,  $g(\lambda^{\star}, w_{c}, r)$ will be a monotonically increasing function of $w_{c}$ for any given $r$.} 

\begin{proof}
	See Appendix \ref{optimal_w}.
\end{proof}

Based on Proposition 2, it is {possible to show} whether the optimization object formulated in (8) or that formulated in (10) will result in the same optimum solution of $w_{c}=2$. More details {can} be found in Appendix C.


\noindent \textbf{Proposition 3.} \emph{{By employing} a regular LDPC parity check matrix, having column weight  $w_{c}$, number of rows $L$, number of columns $N$ and girth larger than 4 as the spreading matrix $\mathbf{S}$, {$R_{\rm FA}(\lambda, w_{c}, r)$ will be a monotonically decreasing function of $r$ for any given $\lambda$ and $w_{c}$.}  }

\begin{proof}
	See Appendix \ref{optimal_r}.
\end{proof}

After substituting $w_{c}=2$ into constraint $\rm C_{3}$, the original optimization object given in (8) can be simplified as (11).
\begin{equation}
\begin{aligned}
\min \limits_{r} \quad &r\\
s.t. \quad &{\rm C_{2}},\\
& R_{\rm FA}(\lambda^{\star},  2, r ) \leq \tau 
\end{aligned}
\end{equation}

Apparently, based on Proposition 3, the optimization problem in (11) is tractable and the optimum solution of $r$, namely $r^{\star}$ is determined by $\tau$.
The value of $r^{\star}$with  several typical $\tau$ are listed in TABLE I.

\begin{table}[htbp]\small
	\centering
	\caption{ $r^{\star}$ with different $\tau$}
	\label{p1}
	\begin{tabular}{ c|c|c|c|c }
		\hline	
		$\tau$ & 0.5 & 1 & 1.5 &2 \\
		\hline
		$r^{\star}$ &0.6666 &0.5 &0.4 &0.31 \\
		\hline
	\end{tabular}
\end{table}


\subsection{Energy Detector}
{Both {the} cover decoder and BP decoder are employed for active user identification. Their implementation requires the knowledge of load states of all time-slots, i.e. $\mathbf{ I }$. In order to simplify the numerical analysis in Section III.B, it is assumed that $\mathbf{ I }$ is {perfectly} known at the receiver side. However, in practice, the estimation of $\mathbf{ I }$ is still necessary. Inspired by \cite{bian2021supporting}, an energy threshold based load state estimator is proposed.}


{In particular}, the entire energy of the received packet on the $l^{th}$ time-slot is given by
\begin{equation}
E_{l} = \sum_{k=1}^{K}|\mathbf{y}[k+ K(l-1)]|^{2} = \sum_{u=1}^{N}a_{u}  \vert \vert\mathbf{x}_{u}\vert\vert_{2}+K\delta^{2}.
\end{equation}
In the estimation process of $\mathbf{ I }$,
if the entire energy received on a time slot exceeds a predefined threshold $\tau_{E}$, this time slot will be marked as loaded, i.e. we have $\mathbf{ I }[l]=1$ if $E_{l}\geq \tau_{E}$. Otherwise, $\mathbf{ I }[l]=0$.


\subsection{{BP algorithm based active user identification}}
As {introduced} in Section II, an extra zero symbol is added to alphabet $\mathcal{X}_{u}$ of {each} user for indicating {its} asleep state. Hence, the BP algorithm has an ability to correct some false alarms that {were made} by {the} cover decoder in the first round of active user identification. The principle is simply that if the probability of $ P\{a_{u}=0\} $ calculated by BP algorithm is sufficiently high, the BP algorithm will assign $a_{u}=0$ and {the $u^{th}$ user is removed}
from the estimated active user set $\hat{ \mathcal{U} }_{ac}^{\mathbf{a} }( \hat{\mathbf{ I }} )$.

In particular, the BP algorithm will iteratively execute (13) and (14) $T$ times and then make its final estimation of $ P\{a_{u}=0\} $.
\begin{equation}
E_{c \rightarrow v}^{(i)} = 
\log(   \frac{1}{  1-   \prod \limits_{v'\in \mathcal{N}(c)\backslash\{v\} } (   1+\exp(   E_{v' \rightarrow c}^{(i-1)}    )    )^{-1}           }        ), 
\end{equation}
\begin{equation}
E_{v \rightarrow c}^{(i)}=
l_{\rm MPA}(a_{v}) + \sum_{c'\in \mathcal{N}(v)\backslash\{c\}}E_{c' \rightarrow v}^{(i-1)},
\end{equation}
where $c$ and $v$ represent check node and variable node of a factor graph, respectively. This factor graph is transformed from the spreading matrix $\mathbf{S}$ that defined in (2). To be {more specific}, only the columns whose indices are included in
$\hat{ \mathcal{U} }_{ac}^{\mathbf{a} }( \hat{\mathbf{ I }} )$
are remained, hence $\mathbf{S}$ is reduced to 
$\mathbf{S}[:,\hat{ \mathcal{U} }_{ac}^{\mathbf{a} }( \hat{\mathbf{ I }} )]$. All zero rows in   $\mathbf{S}[:,\hat{ \mathcal{U} }_{ac}^{\mathbf{a} }( \hat{\mathbf{ I }} )]$ are further removed, {and} resultant matrix could be denoted by $\mathbf{S}_{\rm BP}$. Based on $\mathbf{S}_{\rm BP}$, {the} factor graph is obtained by regarding its rows (i.e. time-slots) as check nodes and its columns (i.e. user activity state) as variable nodes.
Hence, we have $c \in \{ l \vert\mathbf{ I }[l]=1\} $ and  $v \in \hat{ \mathcal{U} }_{ac}^{\mathbf{a} }( \hat{\mathbf{ I }} )$.
Then, $\mathcal{N}(c)$ denotes the set of variable nodes that {are} connected with the check node $c$ on the factor graph. 
$\mathcal{N}(c)\backslash\{v\}$ means remove the variable node $v$ out of $\mathcal{N}(c)$.
Furthermore, $l_{\rm MPA}(a_{v})$ is the  \emph{a priori}  information of the activity state $a_{v}$, which is provided by the data decoding component in Fig. \ref{SIC-MPA}. Moreover, $E_{c \rightarrow v}^{(i)}$ is the extrinsic information transferred from check node $c$ to variable node $v$ within the $i^{th}$ iteration. Particularly, $E_{v \rightarrow c}^{(i)}=0$ for $i \textless 0$, and hence 
$E_{v \rightarrow c}^{(0)}= l_{\rm MPA}(a_{v})$.

Accordingly, at the last iteration, BP algorithm will output the log-likelihood ratio (LLR) of activity state $a_{v}$ as follows
\begin{equation}
l_{\rm BP}(a_{v}) = \log\left( \frac{P(a_{v}=1)}{P(a_{v}=0)}  \right)=l_{\rm MPA}(a_{v}) + \sum_{ c \in \mathcal{N}(v) } E_{c \rightarrow v}^{(T)}. 
\end{equation}

Finally, {the} BP algorithm will regard user $v$ as asleep if { $l_{\rm BP}(a_{v}) \textless 0 $ } and remove it out of previously estimated active user set  $\hat{ \mathcal{U} }_{ac}^{\mathbf{a} }( \hat{\mathbf{ I }} )$. In this case, current factor graph  will be reconstructed {accordingly}.

\section{{Payload} Data Decoding}
{The classical MPA  \cite{hoshyar2008novel} is employed to realize the payload data decoding of all active users. The factor graph required for implementing MPA algorithm is the same as that used in Section III.D, i.e. it can directly leverage $\mathbf{S}_{\rm BP}$. Without loss of generality, let us focus on the decoding of the $k^{th}$ transmitted symbol of every active user. Accordingly, the object of MPA is to estimate the probability of $P(\mathbf{x}_{u}[k] = \mathcal{X}_{u}[m])$ for every $u \in \hat{ \mathcal{U} }_{ac}^{\mathbf{a} }( \hat{\mathbf{ I }} )$ and $m \in [0,M]$, where $\mathbf{x}_{u}[k]$ is defined in Section II, rightly before (1), $\mathcal{X}_{u}[m]$ denotes the $m^{th}$ legitimate symbol in the alphabet of user $u$, particularly, $\mathcal{X}_{u}[0]$ is a zero symbol.
	Let $E_{c \rightarrow v}^{(i)}( \mathcal{X}_{v}[m] )$ denotes
	the extrinsic information of $P(\mathbf{x}_{v}[k] = \mathcal{X}_{v}[m])$ transmitted from check node $c$ to variable node $v$ in the $i^{th}$ MPA iteration.
	Let $E_{v \rightarrow c}^{(i)}( \mathcal{X}_{v}[m] )$ {denote}
	that transmitted from variable node $v$ to check node $c$. Hence, the implementation of our MPA could be summarized as repeats of following equations.}
\begin{equation}
\begin{aligned}
&E_{c\rightarrow v}^{(i)}({  \mathcal{X} }_{v}[m]) =\\
& \sum\limits_{ \mathclap{ m'\in \left[0:M \right] }      \hfill\atop
	\mathclap{v' \in \mathcal{N}(c)\backslash \{ v\}}  }  
\exp(  -\frac{1}{2\delta^{2}}||\mathbf{y}_{k}[c]-\sum\limits_{ \mathclap{ v' \in \mathcal{N}(c)\backslash\{v\} } } \mathbf{S}[c,v']{  \mathcal{X} }_{v'}[m'] -{  \mathcal{X} }_{v}[m]   ||^{2}  )\times\\
&\prod\limits_{\mathclap{v' \in \mathcal{N}(c)\backslash\{v\}}} E_{v'\rightarrow c}^{(i-1)}({  \mathcal{X} }_{v'}[m'])    ,
\end{aligned}
\end{equation}
\begin{equation}
E_{v\rightarrow c}^{(i)}({  \mathcal{X} }_{v}[m]) = \prod\limits_{\mathclap{c'\in \mathcal{N}(v)\backslash\{c\}}} E_{c'\rightarrow v}^{(i-1)}({  \mathcal{X} }_{v}[m]),
\end{equation}
{Since the factor graph used in MPA is the same as that in Section III.D, check node $c$ and variable node $v$ involved in (16-17) will have the same range as that mentioned in Section III.D, i.e. we have $c \in \{l \vert \mathbf{ I }[l]=1\}$ and $v \in \hat{ \mathcal{U} }_{ac}^{\mathbf{a} }( \hat{\mathbf{ I }} )$. According to the same spirit, the notations of $\mathcal{N}(c)$,  $\mathcal{N}(v)$,  $\mathcal{N}(c)\backslash(v)$, $\mathcal{N}(v)\backslash(c)$ remain their definitions specified in Section III.D.}

{At the last iteration of MPA, namely the $T^{th}$ iteration, it will calculate the  \emph{a posteriori} information of 
	every $P(\mathbf{x}_{v}[k] = \mathcal{X}_{v}[m])$ and record them in the probability matrix $\mathbf{P}_{\rm MPA}^{k}$, which could be formulated as}
\begin{equation}
\mathbf{P}_{\rm MPA}^{k}[m,v] = C_{m,v} \prod \limits_{c \in \mathcal{N}(v)} E_{c\rightarrow v}^{(T)}(\mathcal{X}_{v}[m])  ,
\end{equation}
{where the parameter $C_{m,v}$ is invoked for constraining $\mathbf{P}_{\rm MPA}^{k}[m,v] \in [0,1]$.}

As demonstrated in Fig. \ref{SIC-MPA}, the BP algorithm based active user identification and MPA based payload data decoding will iteratively exchange their information. During such an outer iteration, on {the one} hand, MPA will generate $l_{\rm MPA}(a_{v})$ that used in (14) for providing a priori
information to the preceding BP based active user identification. $l_{\rm MPA}(a_{v})$ is calculated according to (19)
\begin{equation}
l_{\rm MPA}(a_{v}) = \log(  \frac{  1-\frac{1}{K}\sum_{k=1}^{K}\mathbf{P}_{\rm MPA}^{k}[0,v]   }{  \frac{1}{K}\sum_{k=1}^{K}\mathbf{P}_{\rm MPA}^{k}[0,v]    }    ).
\end{equation}
{On the other hand, the \emph{a posterior} information of activity state $a_{v}$, namely $l_{\rm BP}(a_{v})$ that {is} obtained from (15), will be input to MPA as its $\emph{a priori}$ information. MPA will utilize $l_{\rm BP}(a_{v})$ to initialize $E_{v \rightarrow c}^{(0)}(\mathcal{X}_{v}[m])$ that {is involved} in (16-17) as follows}
\begin{equation}
E_{v \rightarrow c}^{(0)}\left(\mathcal{X}_{v}[m]\right)= \begin{cases}\frac{1}{1+\exp \left(l_{\mathrm{BP}}\left(a_{v}\right)\right)}, & m=0 \\ \frac{1}{M}\left(1-\frac{1}{1+\exp \left(l_{\mathrm{BP}}\left(a_{v}\right)\right)}\right). & 1 \leq m \leq M\end{cases}
\end{equation}
When the outer iteration between BP algorithm and MPA is terminated, MPA will output its hard decision of $k^{th}$ symbol transmitted from active user $v$. This operation could be formulated as
\begin{equation}
\begin{aligned}
\hat{m} &=\underset{0 \leq m \leq M }{\arg \max } \ \mathbf{P}_{\mathrm{MPA}}^{k}[m, v], \\
\hat{\mathbf{x}}_{v}[k] &=\mathcal{X}_{v}[\hat{m}],
\end{aligned}
\end{equation}

The aforementioned outer iterations between BP algorithm based active user identification and MPA based payload data decoding, as well as the inner iterations within themselves, are summarized as pseudo codes as shown in 
	$\mathbf{{Algorithm 1}}$.

\begin{algorithm}[htbp] 
	\caption{Two-stage iterative detection}
	\label{energy_detector}
	\begin{footnotesize}
		\begin{algorithmic}[1] 
			\Require $\mathbf{y}_{k}$, $k\in [1,K]$, $\delta^{2}$, $\hat{\mathbf{I}}$, $T_{\rm MPA}$, $T_{\rm BP}$, $T_{\rm outer}$;	
			\Ensure $\hat{ \mathcal{U} }_{ac}^{\mathbf{ a }}( \hat{ \mathbf{ I } } )$, $\hat{\mathbf{x}}_{v}[k], v\in \hat{ \mathcal{U} }_{ac}^{\mathbf{ a }}( \hat{ \mathbf{ I } } ),k\in [1,K]$;
			
			\State Estimate  	$\hat{ \mathcal{U} }_{ac}^{\mathbf{ a }}( \hat{ \mathbf{ I } } )$ by cover decoder; 
			\State Generate factor graph based on $\mathbf{S}_{\rm BP}$;
			\For{ $t \in \{  0:T_{\rm outer} \} $ }	
			\If{$t>0$} 
			\State $E_{v \rightarrow c}^{(0)}\left(\mathcal{X}_{v}[m]\right)= \begin{cases}\frac{1}{1+\exp \left(l_{\mathrm{BP}}\left(a_{v}\right)\right)},m=0\\ \frac{1}{M}\left(1-\frac{1}{1+\exp \left(l_{\mathrm{BP}}\left(a_{v}\right)\right)}\right),1 \leq m \leq M\end{cases}$
			\Else
			\State $E_{v \rightarrow c}^{(0)}\left(\mathcal{X}_{v}[m]\right)=\frac{1}{M+1}, \forall     v\in \hat{ \mathcal{U} }_{ac}^{\mathbf{ a }}( \hat{ \mathbf{ I } } ) $	;
			\EndIf
			\For{$k\in \{  1:K \}$}  
			\For{$i\in\{0:T_{\rm MPA}\}$}
			\State Update $E_{c\rightarrow v}^{(i)}({  \mathcal{X} }_{v}[m])$ according to (16);
			\State Update $E_{v\rightarrow c}^{(i)}({  \mathcal{X} }_{v}[m])$ according to (17);
			\EndFor
			\State $\mathbf{P}_{\rm MPA}^{k}[m,v] = C_{m,v} \prod \limits_{c \in \mathcal{N}(v)} E_{c\rightarrow v}^{(T_{\rm MPA})}(\mathcal{X}_{v}[m])  $;
			\EndFor
			\State $l_{\rm MPA}(a_{v}) = \log(  \frac{  1-\frac{1}{K}\sum_{k=1}^{K}\mathbf{P}_{\rm MPA}^{k}[0,v]   }{  \frac{1}{K}\sum_{k=1}^{K}\mathbf{P}_{\rm MPA}^{k}[0,v]    }    )$;
			\For{$i\in \{0:T_{\rm BP}\}$}  
			\State Update $E_{c\rightarrow v}^{(i)}$ according to (13);
			\State Update $E_{v\rightarrow c}^{(i)}$ according to (14);
			\EndFor
			\State $l_{\rm BP}(a_{v}) = l_{\rm MPA}(a_{v}) + \sum \limits_{ c \in \mathcal{N}(v) } E_{c \rightarrow v}^{(T_{\rm BP})}$;
			\For{$ \forall v\in \hat{\mathcal{U}}_{\rm{ac}}^{ }$}
			\If{ $l_{\rm BP}(a_{v})<0$ }
			$\hat{ \mathcal{U} }_{ac}^{\mathbf{ a }}( \hat{ \mathbf{ I } } )  = \hat{ \mathcal{U} }_{ac}^{\mathbf{ a }}( \hat{ \mathbf{ I } } ) - \{v\}$    
			\EndIf
			\EndFor
			\State Update factor graph;
			\EndFor 
			\For{$k\in \{  1:K \}$} 
			\For{$v\in \hat{ \mathcal{U} }_{ac}^{\mathbf{ a }}( \hat{ \mathbf{ I } } )$}
			\State $\hat{m} =\underset{0 \leq m \leq M }{\arg \max } \ \mathbf{P}_{\mathrm{MPA}}^{k}[m, v]$;
			\State $\hat{\mathbf{x}}_{v}[k] =\mathcal{X}_{v}[\hat{m}]$;
			\EndFor
			\EndFor			
		\end{algorithmic}
	\end{footnotesize}
\end{algorithm}

\section{Complexity Analysis}
Considering the proposed two-stage iterative detection that {is} summarized in  $\mathbf{{Algorithm 1}}$, its complexity is imposed mainly by the MPA based payload data decoding component. Then, the {computational complexity} of MPA is further dominated by the calculations that required by every check nodes on the factor graph. Hence, the complexity of the proposed two-stage iterative detection is determined by the degree distribution on $\mathbf{ S }_{\rm BP}$ that introduced below (14). Accordingly, the complexity order of the proposed two-stage iterative detection is given by
\begin{equation}
\mathcal{C}_{} = \mathcal{O}(K\sum_{w=1}^{w_{r}} p_{w}L(M+1)^{w}),
\end{equation}
where $p_{w}$ denotes the proportion of the check nodes, which have a degree of $w$ to all check nodes on the factor graph $\mathbf{ S }_{\rm BP}$.

\noindent \textbf{Proposition 4.} \emph{By employing a regular LDPC parity check matrix having  row weight $w_{r}$,number of rows $L$, number of column $N$ and girth  larger than 4 as the spreading matrix $\mathbf{S}$ {and assuming that} the load states $\mathbf{ I }$ is perfectly estimated,
	$\sum_{w=1}^{w_{r}} p_{w}(M+1)^{w}$  can be approximated as
}
\begin{equation}
\sum_{w=1}^{w_{r}} p_{w}(M+1)^{w} \approx p_{w_{1}}(M+1)^{w_{1}} + p_{w_{2}}(M+1)^{w_{2}},
\end{equation}
where 
\begin{equation}
\begin{aligned}
&p_{w_{1}}=1 - [ \lambda w_{r}- \lfloor \lambda w_{r} \rfloor] \\
&p_{w_{2}}= 1- p_{w_{1}}\\
&w_{1}=   \lfloor \lambda w_{r} \rfloor  \\
&w_{2}= w_{1} +1,
\end{aligned}
\end{equation}
and $\lfloor \cdot \rfloor$ {returns the largest integer that less than or equal to its input.}

\begin{proof}
	See Appendix \ref{complexity_bound}.
\end{proof}
Hence, the complexity {order of the proposed two-stage iterative detection can be approximated as } 
\begin{equation}
\mathcal{C}_{\rm approx} = \mathcal{O}(  KL(  p_{w_{1}}(M+1)^{w_{1}} + p_{w_{2}}(M+1)^{w_{2}} )    ),
\end{equation}
Compared with {the detection method proposed in} \cite{zhu2010exploiting}, whose complexity order is $\mathcal{O}(KL(M+1)^{w_{r}})$, the complexity order of {our proposal} is significantly reduced.
\begin{figure}[htbp]
	\centering
	\includegraphics[scale=0.26]{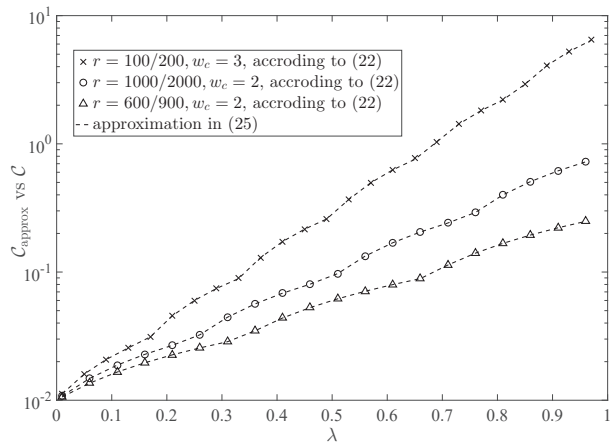}
	\caption{ {Verification of (25)}.  }
	\label{complexity}
\end{figure} 
In order to verify the accuracy of our complexity order approximation, we compare $\mathcal{C}_{\rm approx}$ obtained from (25) with $\mathcal{C}$ obtained from (22) in Fig. \ref{complexity}. In order to guarantee the generality, {different choices for the} spreading matrix $\mathbf{ S }$ and different user sparsity $\lambda$ are tested, which will further result in different factor graph $\mathbf{ S }_{\rm BP}$ for MPA module. The detailed system configuration can be observed at the top left corner of Fig. \ref{complexity}. Numerical results shown in Fig. \ref{complexity} confirm that {replacing} (22) by (25) is reasonable.

\section{Simulation Result and Discussion}

{In this section, the performance of active user identification and payload data decoding are {demonstrated by using computer simulations}. Similar to \cite{jiang2020joint}, when {simulating} the symbol error rate (SER), a symbol is regarded as correct if and only if both the user activity state and the symbol itself are correctly detected. 
	{$\tau_{E}$ mentioned in the Section III.C is fixed at $1.55K\delta^{2}$.}
	{The overloading factor which is also referred to as utilization factor in \cite{abebe2021multi} is defined as $\frac{\lambda N}{L}$ \cite{oyerinde2019compressive}.}
	Then, the system parameters are listed in TABLE II.}

\begin{table}
	\centering
	\caption{Simulation  configuration}
	\label{parameters}
	\begin{tabular}{c | c}
		\hline	Parameter    & value \\
		\hline  {Number of coexisted users} $N$   &$800$ \\
		{Number of time-slots} $L$  &$400$ \\
		{Active user sparsity} $\lambda$  &$0.02, 0.03, 0.04, 0.1, 0.5$\\
		Column Weight $w_{c}$ of $\mathbf{S}$   &$2$\\
		Row Weight $w_{r}$ of $\mathbf{S}$   &$4$\\
		The Value of $\tau$ in (6)      & $1$ \\	
		The Value of $r^{\star}$ in (12)  & $\frac{1}{2}$\\	
		{Constellation size} $M$    & $2$ \\
		{Data Packet Length $K$ } & $60$ \\
		\hline
	\end{tabular}
\end{table}

\begin{figure}[htbp]
	\centering
	\includegraphics[scale=0.26]{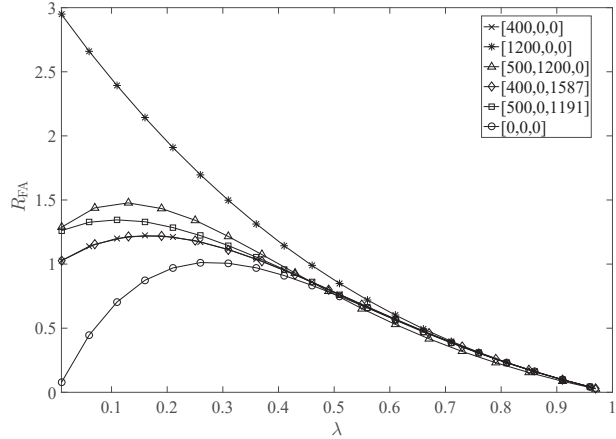}
	\caption{Impact of circle-length distribution of spreading matrix $\mathbf{ S }$ on false alarm performance of cover decoder.}
	\label{short_girth}
\end{figure}

The impact of circle-length distribution of the spreading matrix $\mathbf{ S }$ on the false alarm performance $R_{\rm FA}$ is demonstrated in Fig. \ref{short_girth}, where the number of time-slots, the number of coexisted users, and the thresholds involved in (6) are fixed to $L=400,N=800,\tau=1,r^{\star}=0.5$, respectively. Then, different circle-length distribution could be realized by adjusting the positions of non-zero elements in $\mathbf{S}$. We use the notation of $[400,0,0]$ to represent a circle-length distribution that has 400 length-4 circles, no length-6 circles, and no length-8 circles either. For another instance, $[500, 0, 1191]$ means the constructed spreading matrix $\mathbf{ S }$ has 500 length-4 circles, no length-6 circles, and 1191 length-8 circles. The number of other circles whose length is higher that 8 is not taken into account. By comparing the curve with cross legends to that with star legends, it is clear that the growth of length-4 circles will significantly impair the false alarm performance of the cover decoder component. In contrast, the impact of length-8 circles on $R_{\rm FA}$ is negligible, since the curve with cross legends almost overlap the curve with diamond {marks}, even the later has 1587 more length-8 circles. By comparing the curve with triangle {marks} to that with square {marks}, we may conclude that false alarm performance $R_{\rm FA}$ may slightly degrade owing to having too many length-6 circles. Finally, the curve with circle {marks} confirms that it is better to remove all the short circles in the spreading matrix $\mathbf{ S }$ for improving the cover decoder performance. In this paper, the progressive edge-growth algorithm proposed in \cite{1377521} is employed to optimize the circle-length distribution on the guidance obtained from Fig. \ref{short_girth}.


\begin{figure}[htbp]
	\centering
	\subfigure[]{
		\begin{minipage}{7cm}
			\centering
			\includegraphics[scale=0.21]{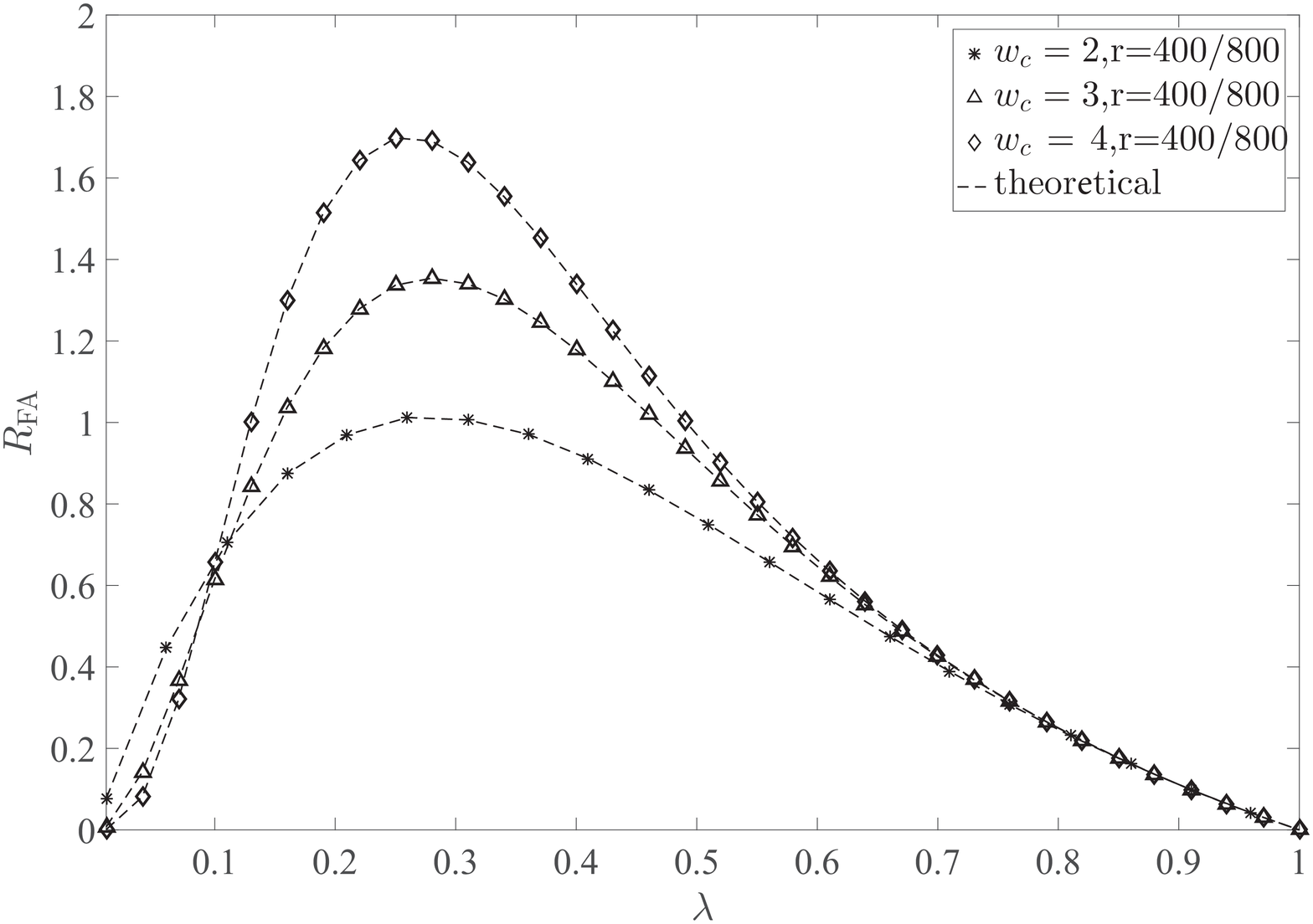}
		\end{minipage}
	}
	\subfigure[]{
		\begin{minipage}{7cm}
			\centering
			\includegraphics[scale=0.21]{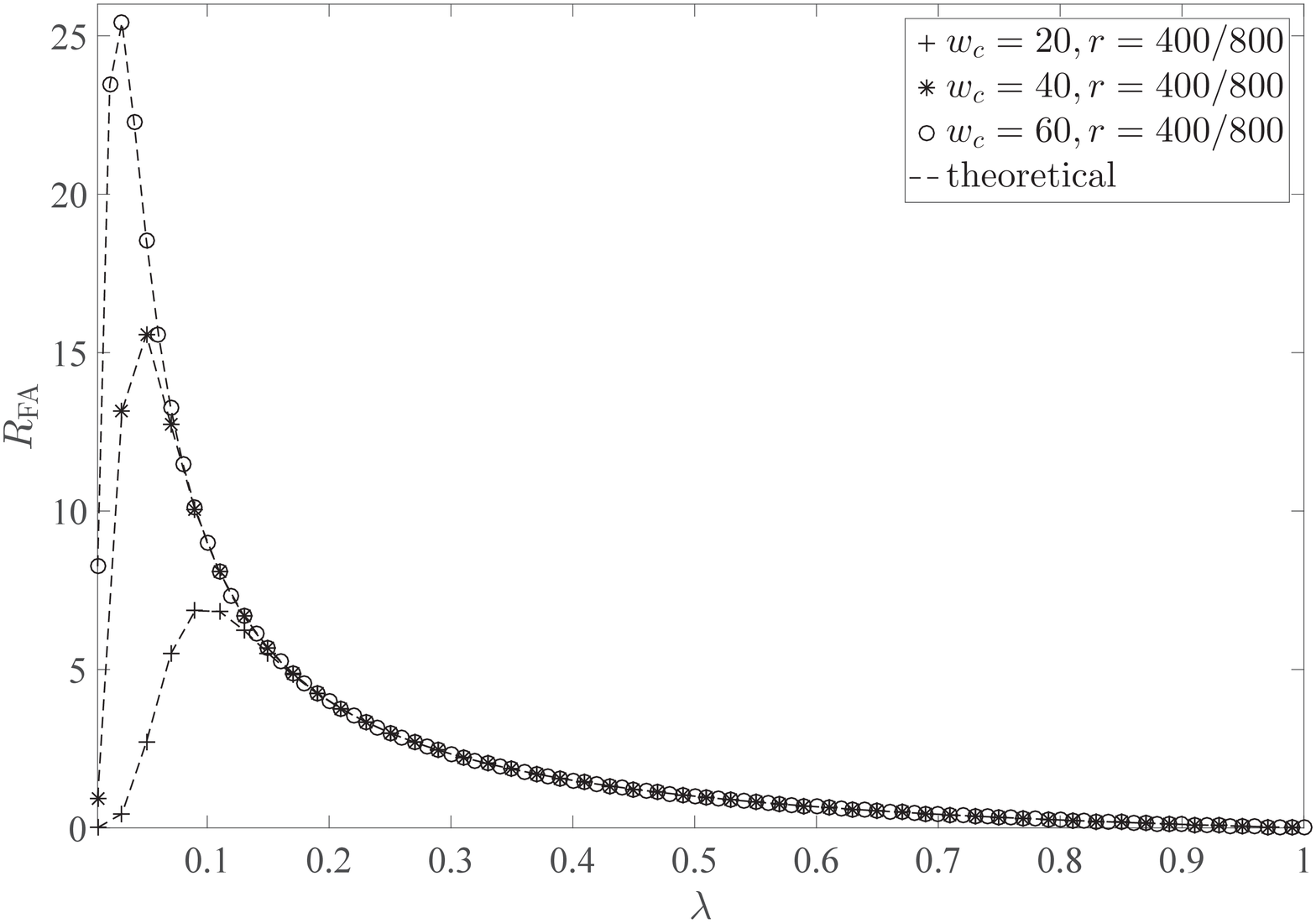}
		\end{minipage}
	}
	\caption{Impact of {column weight $w_{c}$ of spreading matrix $\mathbf{ S }$ on false alarm performance of cover decoder. (a) small $w_{c}$, (b) extremely large $w_{c}$}  }
	\label{wc}
	\vskip -8mm
\end{figure}

Fig. \ref{wc}(a) demonstrates that a smaller {column} weight  $w_{c}$ of spreading matrix $\mathbf{S}$ will contribute a better false alarm performance of cover decoder. This phenomenon confirms the correctness of our \textbf{Proposition 2} that the growth of $w_{c}$ will increase $R_{{\rm FA}}( \lambda^{\star}, w_{c},r )   $.     
The influence of {extremely large} $w_{c}$ is shown in Fig. \ref{wc}(b)
In theory, (7) can not be used to
predict the false alarm performance $R_{\rm  FA}$  when $w_{c} \textgreater 20$.
Because, the assumption adopted in (7) that the girth of the spreading matrix $\mathbf{S}$ {is always} larger than 4 {is no longer satisfied} when $w_{c}$ is {too large, in {fact}, in this case, $\mathbf{S}$ is not a sparse matrix any more. However, the simulation results in Fig. \ref{wc}(b) verify that even the assumptions made in  \textbf{Proposition 2} is not strictly satisfied, (7) still works well and approximates the practical performance. Furthermore, Fig. \ref{wc}(b) confirms again that it is better to employ a small value of $w_{c}$.}



\begin{figure}[htbp]
	\centering
	\includegraphics[scale=0.26]{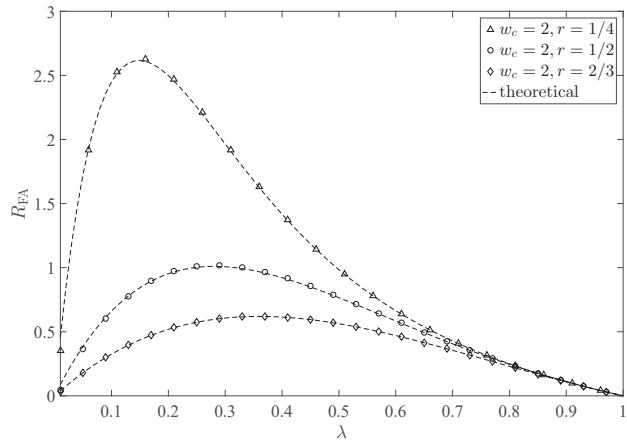}
	\caption{Impact of ratio $r$ of spreading matrix $\mathbf{ S }$ on false alarm performance of cover decoder.}
	\label{r}
\end{figure}
Fig. \ref{r} demonstrates that a larger ratio $r$ of spreading matrix $\mathbf{ S }$ will {lead to} a better false alarm performance of cover decoder. This phenomenon justifies the correctness of our  \textbf{Proposition 3} that the growth of $r$ will decrease  $R_{\rm FA}(\lambda^{\star}, w_{c},r )$.


\begin{figure}[htbp]
	\centering
	\includegraphics[scale=0.26]{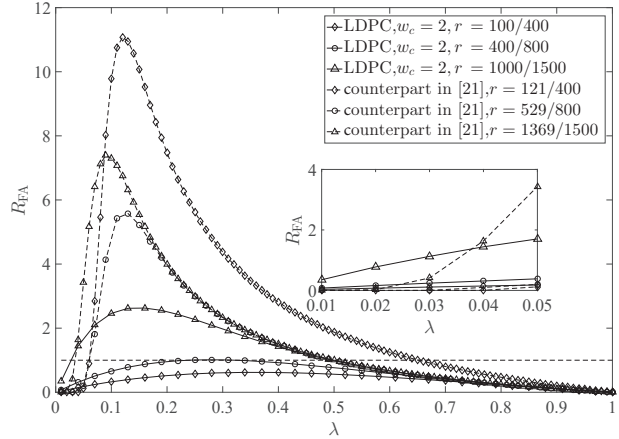}
	\caption{
		False alarm performance $R_{\rm FA}$ comparison between the proposed protocol sequences and  
		its counterpart in \cite{8849823}.}
	\label{LDPC}
\end{figure}
In Fig. \ref{LDPC}, the false alarm performance $R_{\rm FA}$ of our proposal is compared with that of its counterpart in \cite{8849823}.
{As can be seen from Fig. \ref{LDPC},} 
	the false alarm achieved by the proposed protocol sequences significantly outperforms the counterpart in \cite{8849823} as long as $\lambda >0.05$.
	When $0\leq \lambda <0.05$, the false alarm performance $R_{\rm FA}$ of the proposed protocol sequences is only slightly worse than the counterpart in \cite{8849823}.


\begin{figure}[htbp]
	\centering
	\includegraphics[scale=0.26]{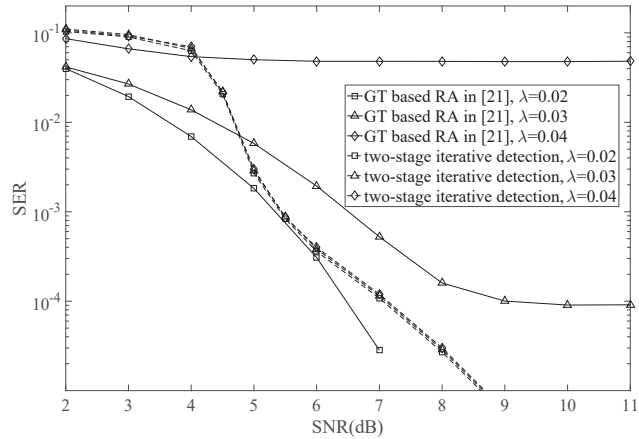}
	\caption{ SER performance comparison between our proposal and its counterpart in \cite{8849823}.}
	\label{counterpart_decoding}
\end{figure}

The SER  comparison between our proposal and the counterpart in \cite{8849823}
is illustrated in Fig. \ref{counterpart_decoding}.
{As can be seen from} Fig. \ref{counterpart_decoding} that the SER performance of our proposal {is significantly better than that of} the counterpart in \cite{8849823} when $\lambda  \geq0.03$.
Furthermore, the SER performance of the counterpart in \cite{8849823} decreases rapidly with the growth of user sparsity $\lambda$. 
In contrast, 
our proposal {is applicable to} a more dense network.
The reason is that the receiver in \cite{8849823} will detect a collision if two or more users are sending packets in the same time-slots and these colliding packets are dropped directly.  In our two-stage iteration detection architecture, these colliding packets are  
possibly recovered by MPA.

\begin{figure}[htbp]
	\centering
	\includegraphics[scale=0.26]{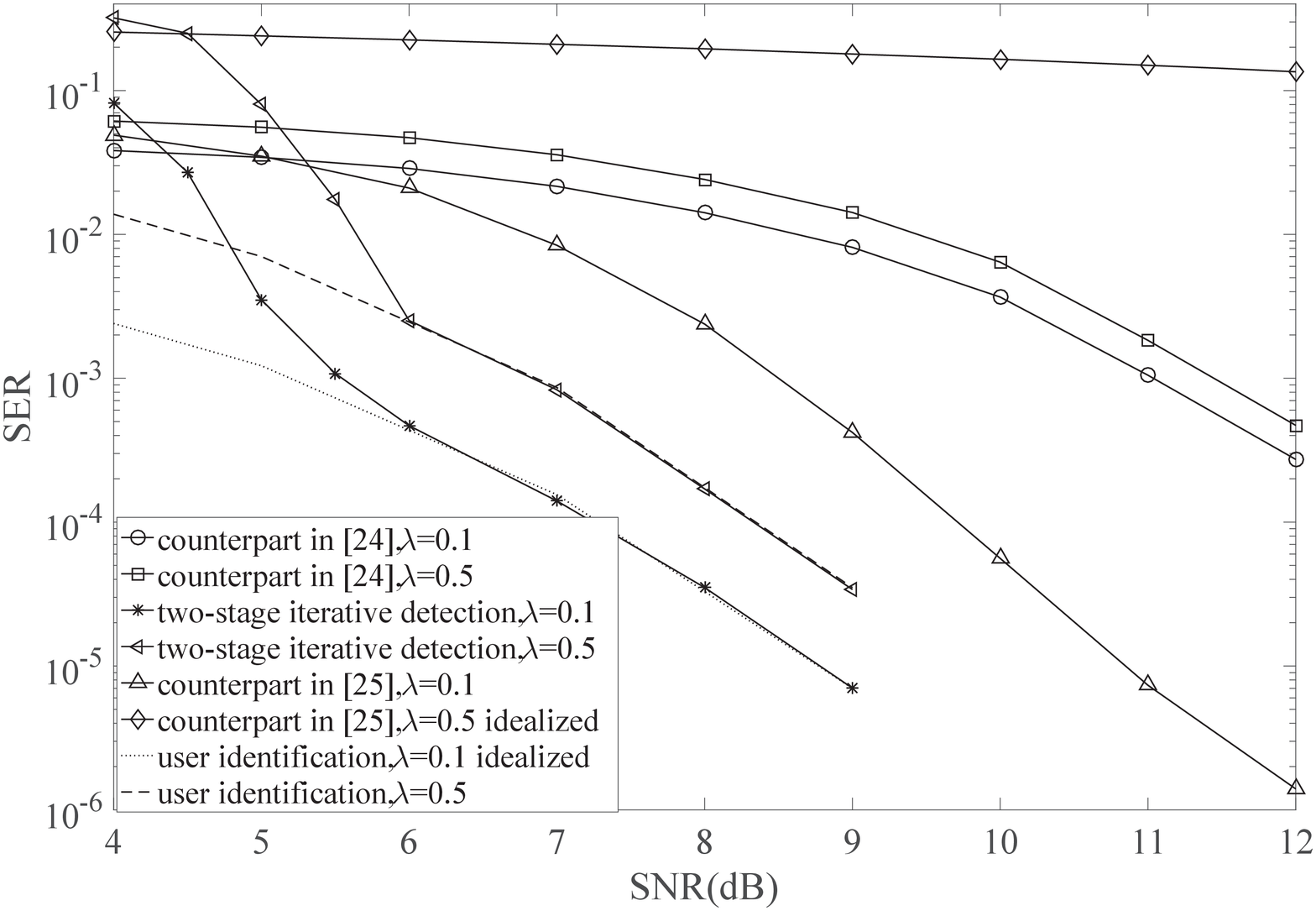}
	\caption{The SER comparison between the counterparts in \cite{zhu2010exploiting,zhang2020bayesian} and the proposed two-stage iteration detection algorithm.}
	\label{SIC-MPA_decoding}
\end{figure}
The SER performance of our proposal in a more dense network
{(e.g. $\lambda = 0.1, 0.5$)}
is shown in Fig. \ref{SIC-MPA_decoding}.
As revealed in Fig. \ref{LDPC}, the performance of counterpart \cite{8849823} will dramatically degrade beyond $\lambda>0.05$ and become significantly inferior to our proposal. Hence, in the dense access model, instead of the counterpart \cite{8849823}, a new detection architecture, namely ``cover decoder plus MPA'' detection \cite{zhu2010exploiting} is further invoked in.
In ``cover decoder plus MPA'' detection, firstly, the user identification is individually completed by a conventional cover decoder with the aid of a perfect estimation of load state $\mathbf{ I }$. {Then}, MPA introduced in Section IV is employed to process data recovery.
Moreover, the counterpart in \cite{zhang2020bayesian} is also considered. 
The proposed Bayesian based message passing algorithm includes two parts, namely Part (i) and Part (ii) respectively. 
With the output message from Part (ii), the active user identification is handled by BP and expectation propagation (EP) algorithm in Part (i). 
In turn, with the incoming message fed back from Part (i), the transmitted data symbols are estimated in Part (ii).
	As can be seen from Fig. \ref{SIC-MPA_decoding} that in the low SNR region of SNR $\leq 5$ dB, the
	 SER performance of the proposed two-stage iterative detection is worse than the counterparts in \cite{zhu2010exploiting} and in \cite{zhang2020bayesian}. 
	 This is due to {the fact that} the proposed energy detector cannot estimate the load states $\mathbf{ I }$ at an acceptable accuracy level during this SNR region.
	In contrast, when $\rm SNR\geq 6$ dB, the proposed two-stage iterative detection significantly outperforms others.
	This phenomenon confirms the advantage of the proposed joint detection between cover decoder and MPA, where the additional outer iterations allow them exchange extrinsic information.
Furthermore, we may assume that the perfect knowledge of load states $\mathbf{ I }$ is always available at the receiver side and hence the practical energy detector employed in Fig. \ref{SIC-MPA} can be removed. The other parts in Fig. \ref{SIC-MPA} are remained.
The resultant receiver is labeled as ``idealized user identification'' in Fig. \ref{SIC-MPA_decoding}. Obviously, it provides a performance upper bound {on} the proposed two-stage iterative detection. It is interesting to find that the proposed two-stage iterative detection {realizes} almost the same performance as that of  ``idealized user identification" beyond $\rm SNR>6$ dB. It implies that the false alarms made by cover decoder can be completely eliminated by the later BP and MPA components as long as the SNR is sufficiently large.


\begin{figure}[htbp]
	\centering
	\includegraphics[scale=0.26]{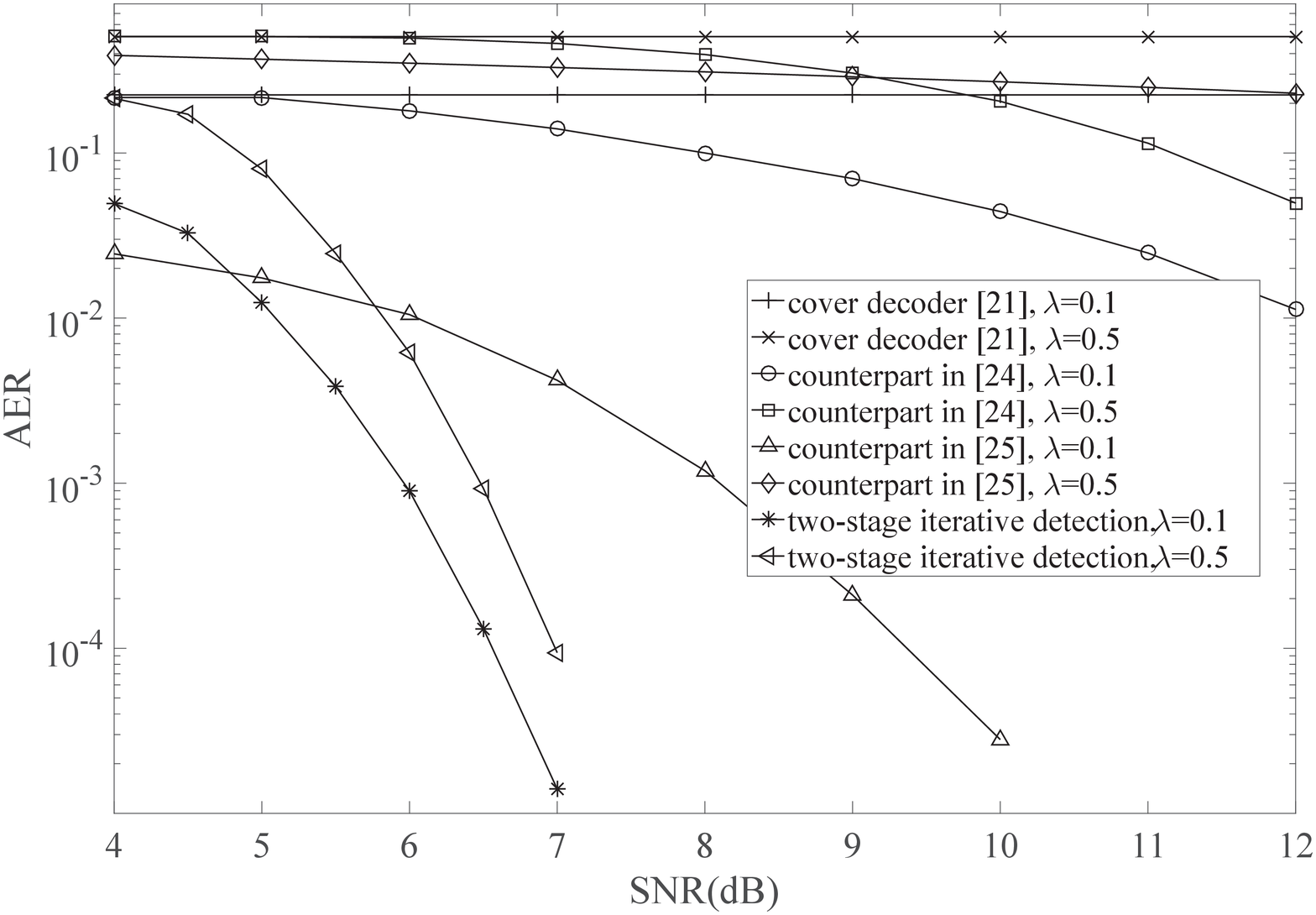}
	\caption{{Active user identification error rate (AER) performance comparison between the proposed two-stage iterative detection and its counterparts in \cite{8849823,zhu2010exploiting,zhang2020bayesian}}}
	\label{AUD}
\end{figure}
The performance of active user identification of the proposed two-stage iterative detection is {shown} in Fig. \ref{AUD}.
	In order to fairly
    measure the active user identification performance, 
    the AER metric proposed in \cite{zhang2020bayesian} is modified to
    $P_{\rm{AER}} = P_{\rm{F}} +P_{\rm{M}}$.    
   , where $ P_{\rm{F}}$ denotes the probability of false detection, $P_{\rm{M}}$ denotes the probability of missed detection.
	Since the counterparts in \cite{8849823,zhu2010exploiting} assume the load states $\mathbf{ I }$ are perfectly known at the receiver,
	the $P_{\rm{M}}$ of their cover decoder is zero. Hence, their AER performances are completely determined by their $ P_{\rm{F}}$.
	As can be seen from
	Fig. \ref{AUD} that our proposal achieves the best AER performance once $\rm SNR > 6 $ dB regardless of user sparsity.
	Then, it is noticeable in Fig. \ref{AUD} that the AER performances of other counterparts in \cite{8849823,zhu2010exploiting,zhang2020bayesian} are constrained by a heavy user sparsity $\lambda$ dramatically .

\begin{figure}[htbp]
	\centering
	\includegraphics[scale=0.26]{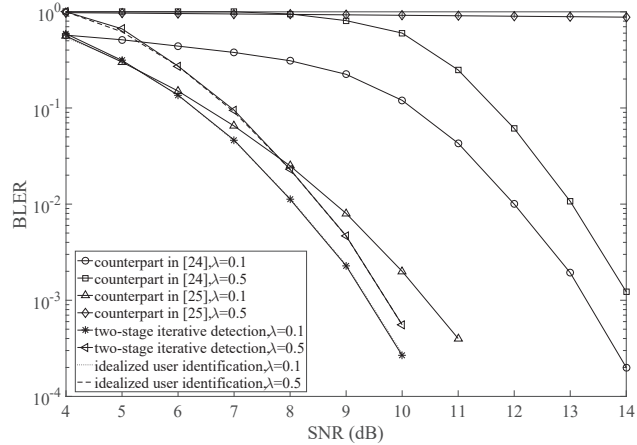}
	\caption{Block error rate (BLER) performance comparison between the proposed two-stage iterative detection and its counterparts in \cite{zhu2010exploiting} and \cite{zhang2020bayesian}.}
	\label{BLER}
\end{figure} 
The BLER performance of the proposed two-stage iterative detection is shown in Fig. \ref{BLER}. 
Again, when the user sparsity $\lambda=0.1$,
	the proposed two-stage iterative detection achieves 
	around 1.2 dB and 4 dB gain at the target of BLER $<10^{-4}$ compared with
	the counterpart in \cite{zhu2010exploiting} and the counterpart in \cite{zhang2020bayesian}, respectively. The performance gain will further grow when user sparsity $\lambda=0.5$.



\section{Conclusion}
In this paper, a novel transceiver architecture {was} designed to achieve efficient random access and reliable data transmission simultaneously.
Explicit preamble transmission {can be} avoided {by} the proposed scheme. Instead, user identities are embedded into channel access patterns and controlled by protocol sequences.
Consequently, the protocol sequence design becomes a critical issue.
Accordingly, the columns of a LDPC parity check matrix are creatively mapped to multiple users' protocol sequences.
Based on the formulation of objective functions for cover decoder, a new parity check matrix construction guideline {was} proposed.
Then, a two-stage iterative detection {was} designed, which leverages group-testing algorithm, BP algorithm and MPA.
Through information transfer between these algorithms, both the active user identification and data recovery are iteratively improved. Our
simulation results show that the false alarms made by the cover decoder can be almost completely corrected by 
the later BP and MPA components in the proposed two-stage iterative detection architecture.
The practical performance of the proposed two-stage iterative detection is capable of approaching the idealized benchmark beyond $\rm SNR \geq 7$ dB.

\begin{appendices}
	\section{}\label{cover_decoder_performance}
	According to equation 
	(24) in \cite{mezard2011group}, 
	when sparsity $\lambda$ is given, the
	average number of false detection of spreading matrix $\mathbf{S}$ whose girth is larger than  $4$ is given by
	\begin{equation}
	\frac{\sum\limits_{ \mathbf{a}\in \mathbb{B}^{N\times 1} \atop ||\mathbf{a}||_{0}=\lambda N }  |\hat{\mathcal{U}}_{\rm{ac}}^{ \mathbf{ a } }(\mathbf{ I })-\mathcal{U}^{  \mathbf{ a } }_{\rm{ac}}|}{    \binom{N}{\lambda N}  } = (1-\lambda) \sum_{u=1}^{N}\prod_{l=1}^{L}(1-(1-\lambda)^{ \frac{w_{c}}{r}-1 })^{\mathbf{S}[l,u]}.
	\end{equation}
	In our design,
	{the} regular LDPC parity check matrix is 
	{employed to construct}
	spreading matrix $\mathbf{S}$. Hence, (26)
	could be further simplified to
	\begin{equation}
	\frac{\sum\limits_{ \mathbf{a}\in\mathbb{B}^{N\times 1}\atop||\mathbf{a}||_{0}=\lambda N}  |\hat{\mathcal{U}}_{\rm{ac}}^{ \mathbf{ a } }(\mathbf{ I })-\mathcal{U}^{ \mathbf{ a }  }_{\rm{ac}}|}{ \binom{N}{\lambda N} }  =(1-\lambda)N(1-(1-\lambda)^{\frac{w_{c}}{r}-1})^{w_{c}}.
	\end{equation}
	Since the size of 
	{${\mathcal{U}}_{\rm{ac}}^{\mathbf{ a }}$}
	is fixed to
	$\lambda N$, the false detection ratio $R_{\rm FA}$ is 
	{evaluated} as
	\begin{equation}
	\begin{aligned}
	R_{\rm FA}(\lambda, w_{c}, r)=&\frac{\sum\limits_{ \mathbf{a}\in \mathbb{B}^{N\times 1}\atop||\mathbf{a}||_{0}=\lambda N }  |\hat{\mathcal{U}}_{\rm{ac}}^{ \mathbf{ a } }(\mathbf{ I })-\mathcal{U}^{ \mathbf{ a }  }_{\rm{ac}}|}{ \sum\limits_{\mathbf{a}\in \mathbb{B}^{N\times 1}\atop||\mathbf{a}||_{0}=\lambda N} \vert \mathcal{U}_{\rm{ac}}^{ \mathbf{ a } }\vert } \\
	=&\frac{(1-\lambda)N(1-(1-\lambda)^{w_{r}-1})^{w_{c}}}{\lambda N}\\
	=& \frac{1-\lambda}{\lambda}(1-(1-\lambda)^{\frac{w_{c}}{r}-1})^{w_{c}},
	\end{aligned}
	\end{equation}

	\section{}\label{optimal_w}
	First, we obtain
	the partial derivative of $g(\lambda, w_{c}, r)$ 
	{with respect to} $w_{c}$,
	which is given by
	\begin{equation}
	\frac{\partial g}{\partial w_{c}} =-[(1-\lambda)^{w_{c}/r-1}+w_{c}(1-\lambda)^{w_{c}/r-1}\ln(1-\lambda)\frac{1}{r}  ],
	\end{equation}
	Let 
	$\frac{\partial g}{\partial w_{c}}\vert_{\lambda=\lambda_{w}}=0$, then the solution is
	$\lambda_{w} = 1- \exp(-\frac{r}{w_{c}})$. 
	Clearly, $\frac{\partial g}{\partial w_{c}}\leq 0$ when $0\leq \lambda \leq \lambda_{w}$ and $\frac{\partial g}{\partial w_{c}}\geq 0$ when $\lambda_{w}\leq \lambda \leq 1$.
	
	Then, we {obtain} the partial derivative of $\ln(R_{\rm FA})$ 
	{with respect to $\lambda$, which is given by}
	\begin{equation}
	\frac{\partial \ln(R_{\rm FA})}{\partial \lambda}=\frac{-1}{\lambda(1-\lambda)} + \frac{ w_{c}( \frac{w_{c}}{r} -1 )(1-\lambda)^{ \frac{w_{c}}{r} -2 }  }{ 1-( 1-\lambda )^{ \frac{w_{c}}{r} -1 }  },
	\end{equation}
	{Substituting} $\lambda_{w}$ to $\lambda$, then (30) becomes
	\begin{equation}
	\begin{aligned}
	\frac{\partial \ln(R_{\rm FA})}{\partial \lambda} \vert_{\lambda=\lambda_{w}} &=  \frac{-1}{e^{-\frac{r}{w_{c}} } (1-\lambda)} + \frac{ w_{c}( \frac{w_{c}}{r} -1 )(  e^{-\frac{r}{w_{c}}  }   )^{ \frac{w_{c}}{r} -2 }  }{ 1-(   e^{-\frac{r}{w_{c}}}   )^{ \frac{w_{c}}{r} -1 }  }\\
	&= \frac{ e^{ \frac{2r}{w_{c}}  }    }{  (e-e^{\frac{r}{w_{c}}})(1-e^{\frac{r}{w_{c}}})   }k(t),
	\end{aligned}
	\end{equation}
	where 
	{$0< t=\frac{r}{w} < 1$}, $k(t)= (e-e^{t}) + w_{c}(\frac{1}{t}-1)(1-e^{t}) $. Let $h(t) = (e-e^{t}) + 2(\frac{1}{t}-1)(1-e^{t})$
	and $k(t) \leq h(t)$.
	
	The derivative of $h(t)$ {can} be written as
	\begin{equation}
	\begin{aligned}
	\frac{{\rm d} h(t)}{{\rm d} t} &= \frac{-2}{t^{2}} + e^{t}(\frac{2}{t^2}-\frac{2}{t}+1)\\
	&= \frac{-2}{t^{2}} + (1+t+\frac{t^2}{2}+\mathcal{O}(t^2))(\frac{2}{t^2}-\frac{2}{t}+1)\\
	&\geq \frac{-2}{t^{2}} + (1+t+\frac{t^2}{2})(\frac{2}{t^2}-\frac{2}{t}+1)\\
	&= \frac{t^{2}}{2} > 0,
	\end{aligned}
	\end{equation}
	Hence, $k(t)<h(1)=0$. 	
	Consequently, $\frac{\partial \ln(R_{\rm FA})}{\partial \lambda} \vert_{\lambda=\lambda_{w}}>0$.
	Hence, $R_{\rm FA}(\lambda, w_{c},r)$ is a concave function of $\lambda$ while fixing 
	$w_{c}$ and $r$. It implies $\lambda_{w} < \lambda^{\star}$.
	
	Finally, 
	$\frac{\partial g}{\partial w_{c}}\vert_{\lambda=\lambda^{\star}}\geq 0$ 
	is obtained. 
	This result reveals
	that the bigger $w_{c}$ is, the bigger $g(\lambda^{\star},w_{c},r)$ is. The proof is accomplished.

	\section{}\label{same_optimal_wc}
	Observe the optimization problem in (10) and (8) that  the constraint $\rm C_{3}$ is guaranteed as long as ${\rm C_{4}}$ is true.
	It implies the set of $w_{c}$ that satisfies (10) must also satisfies (8).
	Let $w_{c, {\rm C_{4}}}^{\star}$ and $w_{c, {\rm C_{3}}}^{\star}$ denotes the optimal $w_{c}$ for
	(10) and for (8), respectively.
	Obviously, $w_{c, {\rm C_{4}}}^{\star}$ is an upper bound {on}  $w_{c, {\rm C_{3}}}^{\star}$, i.e.  $w_{c, {\rm C_{4}}}^{\star}\geq w_{c, {\rm C_{3}}}^{\star}$.
	
	Based on \textbf{Proposition 2}, we 
	{obtain}
	that for any given $\tau$, $w_{c, {\rm C_{4}}}^{\star}$ is always 2. 
	{Owing to the}
	constraint $\rm C_{1}$, $w_{c, {\rm C_{3}}}^{\star}$ is also 2.
	Hence, the fact that the problem in (10) has the same 
	{optimal solution of}
	$w_{c}$ as (8) is proved.

	\section{}\label{optimal_r}
	According to {the} expression of $R_{\rm FA}$, the partial derivative of $R_{\rm FA}$ {with respect to} $r$ {is given by}
	\begin{equation}
	\frac{\partial R_{\rm FA}}{\partial r} = \frac{ w_{c}^{2} (1-\lambda)^{w_{c}/r}\log(1-\lambda) }{\lambda r^{2}} (1-(1-\lambda)^{w_{c}/r-1}  )^{w_{c}-1}.
	\end{equation}
	Obviously, $(1-(1-\lambda)^{w_{c}/r-1}  ) \ge 0$, $\log(1-\lambda) \leq 0$. Hence, $\frac{\partial R_{\rm FA}}{\partial r}\leq 0$. It implies that the bigger $r$ is, the smaller $R_{\rm FA}$ is. The proof is accomplished.

	\section{}\label{complexity_bound}
	{Let} $\mathcal{S}_{ u}^{l}$ denote the {user set who possibly interferes} with user $u$ on the $l^{th}$ time slot, i.e. $\mathbf{S}[l,u']=1, 1\leq u'\neq u \leq N$.
	Since the girth length of $\mathbf{S}$ is {larger than} or equal to 6, 
	{we have}
	$\mathcal{S}_{ u}^{l_{1}}  \cap \mathcal{S}_{ u}^{l_{2}} = \emptyset$, {while} $l_{1} \neq l_{2}$, 
	$\mathbf{S}[l_{1},u] = \mathbf{S}[l_{2},u]=1 $. 
	Then, let random variable $W_{u}^{l}$ {denote} the size of $\mathcal{S}_{ u}^{l}$.
	According to \cite{mezard2011group}, the values of the random variables $W_{u}^{l}, 1\leq l \leq L, \mathbf{S}[l, u]=1$ are independent.
	
	Obviously, the degree of any  check node in the factor graph $\mathbf{S}[:, \mathcal{U}_{ac}^{\mathbf{ a }}]$ is dominated by {size} of 
	the set $ \mathcal{S}_{ u}^{l}\cup \{u\} ,  \mathbf{S}[l,u]=1$.
	Hence, the expectation of the degree of the $l^{th}$ check node $d_{c}^{l}$ can be {formulated} as 
	\begin{equation}
	E[d_{c}^{l}] = E[W_{u}^{l}   ] + E[ a_{u} ].
	\end{equation}

	In our {scenario}, {any column} in the spreading matrix $\mathbf{S}$ has a probability $\lambda$ to be {added} into the spreading matrix of active user set $\mathbf{S}[:, \mathcal{U}_{ac}^{\mathbf{ a }}]$. 
	Hence,  $W_{u}^{l}$ obeys {to} a binomial distribution $B(w_{r}-1, \lambda)$. Hence, $E[W_{u}^{l}   ]= \lambda(w_{r}-1)$. 
	{Furthermore, we have $ E[a_{u}  ]= \lambda$. According to (34),} $E[d_{c}^{l}] = \lambda w_{r}$.
	Moreover, the values  of  $d_{c}^{l}$ must be 
	{an integer}.
	Hence,  {the value} of  $d_{c}^{l}$ 
	{could be a random integer}
	whose value is
	$w_{1}=\lfloor \lambda w_{r}  \rfloor$ with 
	{a probability of}
	$p_{w_{1}}= 1- (\lambda w_{r}- \lfloor \lambda w_{r}  \rfloor)$ and $w_{2} = w_{1}+1$ with 
	{a probability of}
	$1- p_{w_{1}}$.
	The proof is accomplished.

\end{appendices}

\bibliographystyle{IEEEtran}
\bibliography{IEEEabrv,ForIEEEBib}

\end{document}